\documentclass[a4paper]{jpconf}
\usepackage{graphicx}
\usepackage{color}
\usepackage{multirow}

\usepackage[applemac]{inputenc}
\usepackage[english]{babel}

\begin{document}

\title{Propagation of collective modes in non-overlapping dipolar Bose-Einstein Condensates}
\author{A. Gallem\'{\i}, M. Guilleumas, R. Mayol and  M. Pi}
\address{Departament d'Estructura i Constituents de la Mat\`eria,
Facultat de F\'isica, Universitat de Barcelona, 08028-Barcelona, Spain\\
Institut de Nanoci\`encia i Nanotecnologia (IN$^2$UB), Universitat de Barcelona, 08028-Barcelona, Spain}
\ead{gallemi@ecm.ub.edu}

%%%%%%%%%%%%%%%%%%%%%%%%%%%%%%%%%%%%%
%%%  ABSTRACT                                                                          %%%
%%%%%%%%%%%%%%%%%%%%%%%%%%%%%%%%%%%%%

\begin{abstract}
We investigate long-range effects of the dipolar interaction in Bose-Einstein condensates
by solving the time-dependent 3D Gross-Pitaevskii equation.
We study the propagation of excitations between non-overlapping condensates 
when a collective mode is excited in one of the condensates. We obtain 
the frequency shifts due to the long-range character of the dipolar coupling for the bilayer and 
also the trilayer system when the dipolar mode is excited in one condensate.
The propagation of the monopolar and quadrupolar modes are also investigated.
The coupled-pendulum model is proposed to qualitatively explain the long range effects of the dipolar
coupling.
\end{abstract}

% \textbf{Keywords:} Bose-Einstein Condensate, Dipolar interaction, Excitation, Long-range

\section{Introduction}  
\vskip5mm
Recent experiments on dipolar Bose-Einstein Condensates (dBECs) of chromium~\cite{Griesmaier2005}, 
dysprosium~\cite{Lu2011}, and erbium atoms \cite{Aikawa2012}, have stimulated
a growing interest in ultracold dipolar quantum gases 
\cite{Lahaye2009} due to the long-range and anisotropic nature of such interaction. The research on this 
kind of interaction is expecting to give rise to important new features not only at microscopic level, but at macroscopic 
level as well. 
The development of new experimental techniques focused on the realization of molecular gases 
with large electric polar moment, where the dipolar effects are particularly strong, has led to the hope that challenging new
 frontiers in this field can be opened.

The theoretical research in dipolar gases has tended to focus mostly on 
the study of the anisotropic effects of the dipolar interaction \cite{Lahaye2009,Abad2010}, rather than on the long-range 
ones, where there are only few proposals: cat states in triple-well potentials \cite{Lahaye2010,Yan2012} and propagation 
of center-of-mass modes studied by means of a variational method \cite{Matveeva2011,Huang2010}.

The effect of the dipole-dipole interaction on the frequency of a collective mode has been experimentally measured
 in a single dBEC \cite{Bismut2010}. However, a central issue that remains opened is to 
%investigate
understand
the mechanism 
of the expansion and propagation of collective modes between non-overlapping condensates.

The aim of the present work is to study the propagation of dipolar, monopolar and quadrupolar modes induced by the
 long-range nature of the dipolar interaction by solving the full 3D time-dependent Gross-Pitaevskii equation 
(TDGPE). We consider two dipolar Bose-Einstein condensates harmonically trapped in a double-well 
configuration such that the overlap between the two clouds is negligible as well as the corresponding tunneling
 effect. The only effective force acting between the two dBECs is the one produced by the long-range behaviour 
of the dipolar interaction. 

 First, we study the propagation of center-of-mass 
(dipolar) modes between two non-overlapping pancake-shaped condensates.
Our time-dependent numerical results are in agreement with previous ones obtained within the variational method 
\cite{Matveeva2011,Huang2010}. Then, we investigate the propagation of other excitations (monopolar and quadrupolar modes), 
in the bilayer system. That is, the appearance and shift of new characteristic frequencies. 
Finally, as an extension, we test our model in a triple-well configuration, comparing
the results with the case of the double-well configuration, and finding out that the classical model gives a qualitative description
of the dipolar coupling.
The coupled-pendulum model is used to understand qualitatively the results obtained before.
 
%\section{Dipolar interaction in BECs and double-well configuration}
\section{The system: dipolar interaction in BECs and double-well configuration}
\vskip5mm
Bose-Einstein condensates at low temperatures can be described in the mean-field framework 
provided a large number of particles and a weakly interacting system.
The TDGPE in presence of the dipolar interaction reads:
%
%\begin{eqnarray}
% \hbar \frac{\partial\Psi(\vec r,t)}{\partial t}&=&\left[-\frac{\hbar^2}{2m}\nabla^2+
%V_{ext}(\vec r)+g|\Psi(\vec r,t)|^2\right.\nonumber\\
%&&\left.+  d^2 \int{d \vec r\,'|\Psi(\vec r\,',t)|^2\frac{1-3 \cos\theta}{|\vec r-\vec r\,'|^3}}\right]\Psi(\vec r,t)\,.
%\label{diptdgp}
%\end{eqnarray}
%
\begin{equation}
i\hbar \frac{\partial\Psi(\vec r,t)}{\partial t}=\left[-\frac{\hbar^2}{2m}\nabla^2+
V_{ext}(\vec r)+g|\Psi(\vec r,t)|^2+V_D(\vec r) \right] \Psi(\vec r,t)\,,
\label{tdgp}
\end{equation}
where $\Psi(\vec r,t)$ is the wavefunction of the condensate, which is normalized to the number of particles 
$N=\int|\Psi(\vec r,t)|^2 d\vec r$, $m$ is the mass of the bosons, and $V_{ext}(\vec r)$ is the trapping potential.
The contact interaction potential is characterized by the coupling constant
$g=4\pi \hbar^2 a/m$, where
$a$ is the $s$-wave scattering length. The mean-field dipolar interaction is given by
\begin{equation}
 V_D(|\vec r-\vec r\,'|,\theta)=d^2 \,\frac{1-3\cos^2\theta}{|\vec r-\vec r\,'|^3}\,.
\label{vdip}
\end{equation}
Here $\vec d=\sqrt{\mu_0/4 \pi} \, \vec \mu _m$, where $\mu_0$ is the  magnetic permeability of vacuum
and $\vec \mu_m$ is the magnetic dipole moment of the atoms.
All the atoms are assumed to have the same 
dipole moment, both in magnitude and orientation given a polarization direction. 
%Here $\vec d=\sqrt{\mu_0/4\pi} \vec \mu$ is the dipole moment, $\mu_0$  
%is the  magnetic permeability of vacuum, 
The angle $\theta$ is the angle between 
%$\vec d$ 
the magnetization axis
and the relative position between two dipols. 
We can see from Eq.~(\ref{vdip}) that  in contrast to contact interaction,
the dipolar interaction
 is anisotropic, due to the term $\cos^2\theta$, and long-range, due to the $1/|\vec r-\vec r\,'|^3$ dependence.

\begin{figure}[h!]
\begin{center}
\includegraphics[width=0.5\linewidth, clip=true]{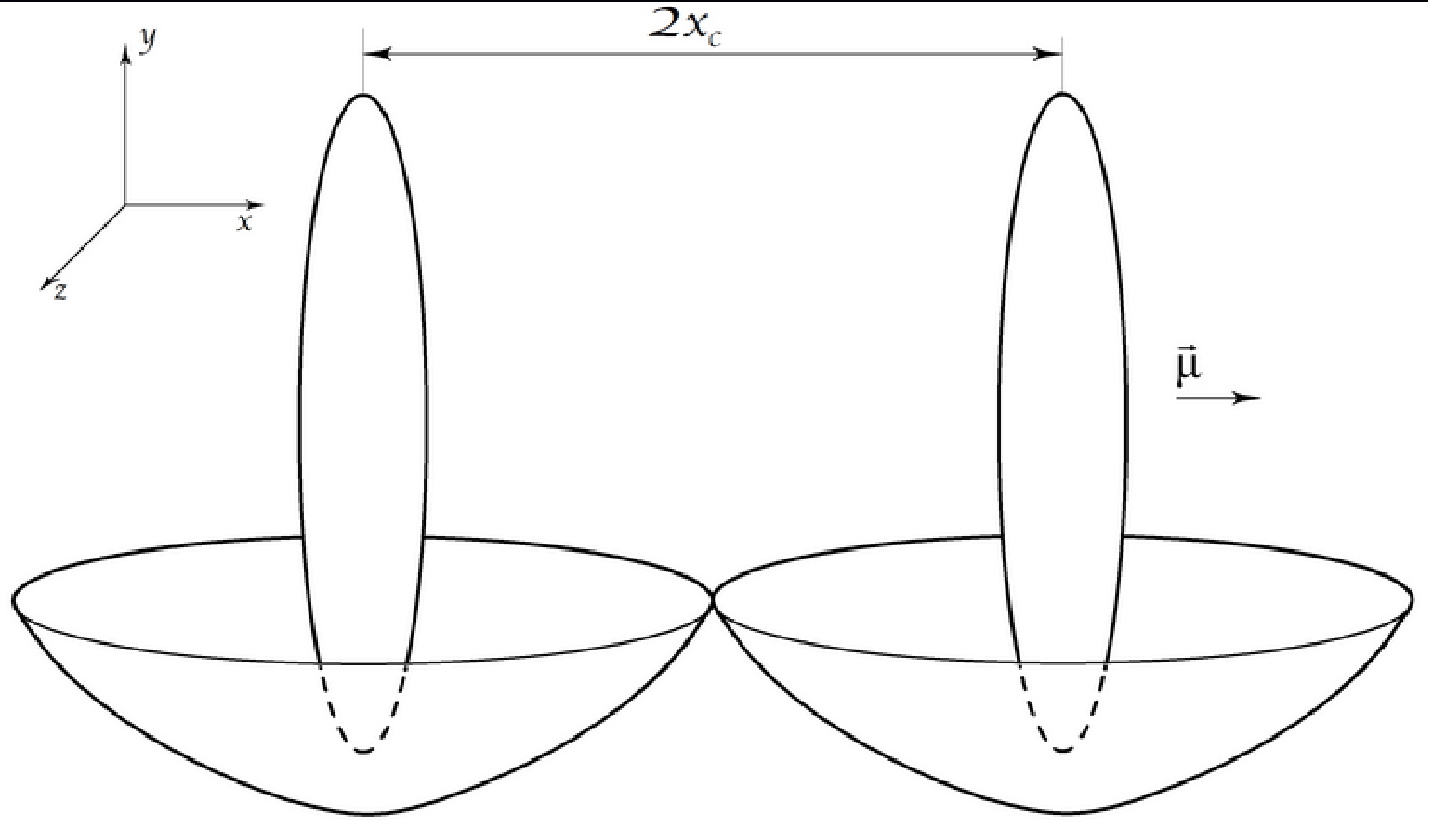}
\end{center}
\caption{\label{doublewell}Illustration of the double-well configuration. The dipole moments are oriented in the $x$-direction, perpendicular to the pancake-shaped dBECs.}
\end{figure}

%The dipolar interaction term is added in the TDGPE in the following way:
%\begin{eqnarray}
%\hbar \frac{\partial\Psi(\vec r,t)}{\partial t}&=&\left[-\frac{\hbar^2}{2m}\nabla^2+V_{ext}(\vec r)+
%g|\Psi(\vec r,t)|^2\right.\nonumber\\
%&&\left.+  d^2 \int{d \vec r\,'|\Psi(\vec r\,',t)|^2\frac{1-3 \cos\theta}{|\vec r-\vec r\,'|^3}}\right]\Psi(\vec r,t)\,.
%\label{diptdgp}
%\end{eqnarray}
% 
In this work, we consider a bilayer system confined in the following
double well potential:
\begin{equation}
 V_{ext}(x,y,z)=\frac{1}{2}m\left(\omega_\perp^2 \, (y^2+z^2)+\omega_x^2\, min[(x+x_c)^2,(x-x_c)^2]\right)\,.
\label{vtrap}
\end{equation}
The shape of each single dBEC depends on the trapping frequencies $\omega_\perp$ and $\omega_x$. 
The distance between the two dBECs is given by $L=2\,x_c$.
In the numerical calculation we have used the following parameters: $2\,x_c=3\mu$m, 
$m=52\;\mbox{uma}$ ($^{52}$Cr atoms) , total number of particles $N=5000$ ($2500$ bosons in each dBEC), and
trapping frequencies
$\omega_x=1500\times2\pi\, \,\mbox{Hz}\gg\omega_\perp=50\times2\pi\,\, \mbox{Hz}$. Thus, 
each dBEC has a pancake-shaped geometry (Fig. \ref{doublewell}).  The oscillator length in the $x$ direction that gives a longitudinal length scale is:
$l_x=\sqrt{\hbar/m \omega_x^2}= 23.21 \mbox{nm}$. The scattering length choosed for the particles is $a=0.001\;a_B$, where $a_B$ is the Bohr 
radius.

The two condensates 
are non-overlapping, that is, the barrier height is large enough  to avoid the tunneling of particles.
But they must be close enough in order to analyze the long-range properties of the interaction
since it decays as $1/|\vec r-\vec r\,'|^3$.
 Finally, the dipole moments are aligned in the $x$-direction, parallel to the symmetry axis of the pancakes.
 This corresponds to the attractive configuration (the repulsive configuration would imply the dipoles to be oriented 
 in the plane of the pancake shaped condensate).
The purpose of this choice is to ensure, first, the stability of the condensate \cite{Lahaye2009} 
(cigar-shaped geometry brings earlier to the collapse than the pancake-shaped), and second, other dipole moment 
orientations would decrease dipolar effects.

%%%%%%%%%%%%%%%%%%%%%%%%%%%%%%%%%%%%%%%%%%%%%
%%% SECTION:   Coupled-pendulum model                %%%
%%%%%%%%%%%%%%%%%%%%%%%%%%%%%%%%%%%%%%%%%%%%% 

\section{The coupled-pendulum model}
\vskip5mm
Before presenting the numerical simulations, a qualitative description of the propagation of a collective 
mode between two non-overlapping dBECs can be found in the frame of classical mechanics 
\cite{Matveeva2011}. We consider the classical analogy of two-coupled pendulums, where the natural 
oscillation frequency of each pendulum ($\omega_0$) corresponds to the frequency of the excited mode, and the 
coupling string $\alpha$ corresponds to the dipolar coupling. 
The two natural modes of oscillation are:
\begin{eqnarray}
\omega_{in}  &=&\omega_0 \label{omegain}                                                     \\
\omega_{out}&=&\omega_0 \sqrt{1+2\alpha/\omega_0^2}\,,\label{omegaout}
\end{eqnarray}
which are the \textit{in-phase mode} and the \textit{out-of-phase mode}, respectively. In the dipolar drag 
effect when the center-of-mass mode is excited, $\omega_{in}$ is the frequency of the trap in the 
direction of the initial displacement (Kohn's theorem).

The parameter $\alpha$, depends on the dipole moment $\vec \mu_m$, the distance between the two condensates 
$2\,x_c$ and their density gradient. Whereas the first two parameters are easily tunable, the latter is more 
difficult to control (in spite of the fact that it can be changed by varying all the parameters). Therefore,
 these two parameters and the frequency of the excited mode will detune the shift of the
 \textit{out-of-phase mode} frequency $\omega_{out}$ \cite{Matveeva2011}.

The previous result can be generalized to $n$ coupled pendulums. The eigenvalues of the problem are:
\begin{equation}
 \omega_{ev}=\omega_0\sqrt{1+(2+\gamma)\alpha/\omega_0^2}\,.
\label{npend1}
\end{equation}
And the parameter $\gamma$ can be found by solving the following equation \cite{Hu1996}:
\begin{equation}
 (1+\gamma)^2\xi(\gamma,n-2)-2(1+\gamma)\xi(\gamma,n-3)+\xi(\gamma,n-4)=0\,,
\label{npend2}
\end{equation}
where
\begin{equation}
 \xi(\gamma,n)=(-1)^n\, \frac{\sinh((n+1)\Omega(\gamma))}{\sinh{\Omega(\gamma)}}\,,\;\mbox{with} 
\;\;\;\Omega(\gamma)=\mbox{arccosh}(-\frac{\gamma}{2})\,.
\label{npend3}
\end{equation}
%

%%%%%%%%%%%%%%%%%%%%%%%%%%%%%%%%%%%%%%%%%%%%%
%%% SECTION: Dipolar Drag Effect                                                                     %%%
%%%%%%%%%%%%%%%%%%%%%%%%%%%%%%%%%%%%%%%%%%%%%

%\section{Dipolar drag effect}
\section{Numerical results}
\vskip5mm
A nice long range effect that can be realized in a bilayer configuration is the dipolar drag.
It is well known, that when the center of mass mode (dipole mode) is excited in a single BEC confined
in a harmonic trap  (by displacing the condensate out of its equilibrium position and releasing it, 
or giving some velocity to the center of mass of the condensate), the condensate oscillates 
harmonically with the same frequency as the characteristic trap frequency in the direction of the 
initial perturbation. This is still true in presence of long-range interactions 
that is for a dipolar Bose-Einstein condensate in a harmonic potential.
However, when the dipole mode is excited in one dBEC of a bilayer system, although 
there is no tunneling between both condensates, the center of mass of the other dBEC starts
to oscillate, as a result of the long-range character of the dipolar interaction,
and both the {\it in-phase} and the {\it out-of-phase} dipole modes of the bilayer system are excited. 
Whereas the {\it in-phase mode} (center of mass) is independent of the
dipolar interaction, the frequency of the {\it out-of-phase} mode depends on it.
Dipolar drag is one of the so called propagation of collective
modes, produced by the long-range nature of the dipolar interaction. It has been already studied in 
electron gases in uniform bilayer systems, under the name of Coulomb drag \cite{Jauho1993,Rojo1999}.

%%%%%%%%%%%%%%%%%%%%%%%%%%%%%%%%%%%%%%%%%%%%%
%%% SECTION: Numerical results                                                                           %%
%%%%%%%%%%%%%%%%%%%%%%%%%%%%%%%%%%%%%%%%%%%%%

\subsection{Dipolar drag effect}
\vskip5mm
In order to observe the dipolar drag effect one has to  take into account the following items:
\begin{itemize}
\item The perturbation that produces the dipolar excitation must be in a transversal direction of the pancake-shaped dBEC, 
because
the confinement in the longitudinal direction ($\omega_x$) is so tight that leads to a small frequency 
shift, see Eq.~(\ref{omegaout}). Here we excite the center of mass mode of the left condensate along $y$-direction.

\item The dipole orientation shall be in the $x$-direction (attractive configuration between the two layers), 
otherwise, the dipolar interaction favors the 
interaction between the atoms of the same condensate rather than the coupling with the other condensate.
\end{itemize}

%\item 
%**** ALBERT, N'ESTEM SEGURS??? Si és un efecte long-range pur li ha de ser igual que hi hagi interacci\'o
%de contacte ****
%
%The scattering length has to be small to ensure that the dBEC is dominated by the dipolar 
%interaction. In this work, $a=0.001\;a_B$, where $a_B=5.2918 \times 10^{-11} \; \mbox{m}$ is the Bohr 
%radius. It is important to remind that  the scattering length can be experimentally tuned by means of 
%Feshbach resonances.

\begin{figure}[h!]
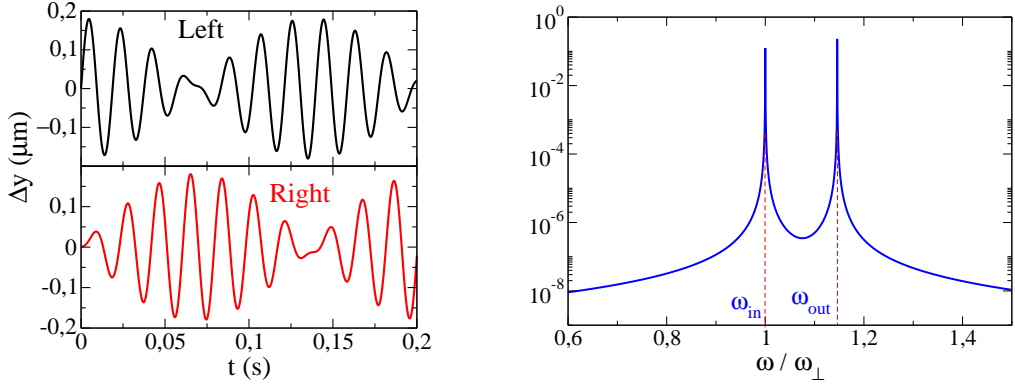

\begin{center}
\includegraphics[width=0.35\linewidth, clip=true]{centremasses2.eps}
\hspace{1cm}
\includegraphics[width=0.4\linewidth, clip=true]{FFT2.eps}
\end{center}
\caption{\label{centremasses}
Left: Diplacement of the center of mass of the left (top) and right (bottom) condensate with the value 
of $d/\mu_B=10$, where $\mu_B$ is the Bohr magneton. Right: Fourier 
transform (logarithmic scale) of the displacement of the center of mass of both condensates. The first peak 
corresponds to $\omega_{in}=\omega_{\perp}=50 \,\mbox{Hz}\times2\pi=\,314.16\,\mbox{Hz}$, whereas the second one is 
$\omega_{out}= 1.14 \, \omega_{\perp}=358.14 \,\mbox{Hz}$.
%$\omega_{out}=\omega_\perp\sqrt{1+2\alpha/\omega_\perp^2}$. 
The strength is plotted in arbitrary units.}
\end{figure}

Taking into account the above considerations we have studied the dipolar drag effect by solving the TDGPE.
The numerical procedure has three steps:

%*****************************************************

\begin{enumerate}
 \item By using the imaginary time step method, we find the stationary solution to reach the ground state of 
the system, which has the form $\Psi(\vec r,t)=\Psi_{GS}(\vec r)  \,exp(-i\epsilon t/\hbar)$, where $\epsilon$ is the 
chemical potential.
 \item Then, we add a momentum to the wavefunction of the left condensate displacing it out of its ground 
state, with the transformation (in the region $x<0$):
\begin{equation}
 \Psi(\vec r, 0) \to \Psi_{GS}(\vec r) \;e^{i\lambda\, y}\,,
\label{dipkick}
\end{equation}
where $\lambda$ quantifies the magnitude of the perturbation. We remind that the kick must be in a 
transversal direction of the pancake, where the BEC is less confined.
 \item The excited new wavefunction is evolved in time by solving Eq.(\ref{tdgp}) using Runge-Kutta and Hamming's 
method (predictor-modifier-corrector) \cite{Abad2011}.
\end{enumerate}

In Fig.$\,$\ref{centremasses} we show the displacements of the centers of mass of both condensates and the 
corresponding Fourier analysis \cite{Calvayrac1997}. One can see that the right dBEC starts to oscillate due to the dipolar coupling. 
The beating of the oscillations comes from the presence of the two normal modes of the coupled system. 
The frequencies are obtained by Fourier analysis, the lower frequency corresponds to the 
\textit{in-phase mode} (\ref{omegain}) and the second peak is the \textit{out-of-phase mode} 
which is slightly shifted from the excitation frequency $\omega_\perp$ (\ref{omegaout}).

\begin{figure}[h!]
\begin{center}
\includegraphics[width=0.6\linewidth, clip=true]{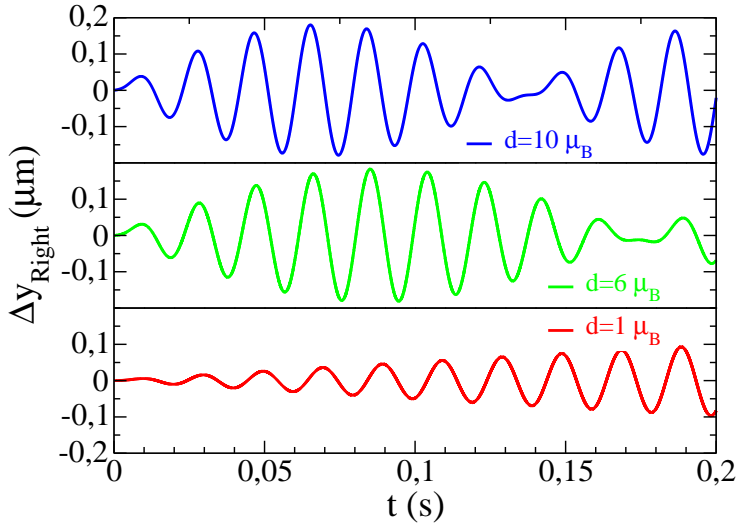}
\end{center}
\caption{\label{compard}
Displacement of the center of mass of the condensate of the right for different dipole moments. From 
top to bottom, $d=10\,\mu_B$, $d=6\,\mu_B$ and $d=1\,\mu_B$.}
\end{figure}

\begin{figure}[h!]
\begin{center}
\includegraphics[width=0.6\linewidth, clip=true]{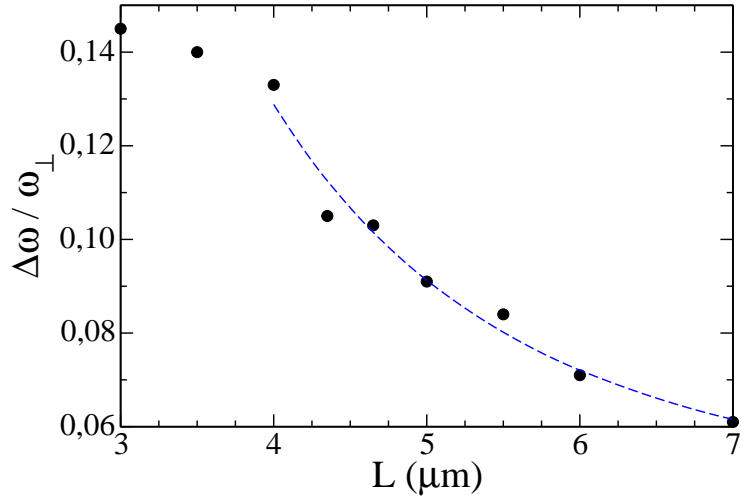}
\end{center}
\caption{\label{distance}
Frequency shift of the center of mass mode as a function of the relative distance of the bilayer system. The line shows the potential decay at large distances.}
\end{figure}

We can analyse now the dependence of the frequency shift on the dipole moment and the distance between the condensates. On the one hand, in Fig. \ref{compard}, we show the displacement of the center of mass of the right dBEC for different 
values of $d$: the frequency shift is larger and the time response is smaller when the dipole moment 
increases. We have used the values: $d/\mu_B=1;\,6 \mbox{ and } 10$. On the other hand, in Fig. \ref{distance}, we plot the frequency shift for several values of $L$, showing the decreasing of the shift when the dBECs get away. When they approach, a change of the potential behaviour occurs because they start to see the structure of dipols of the pancake-shaped condensates.

%%%%%%%%%%%%%%%%%%%%%%%%%%%%%%%%%%%%%%%%%%%%%
%%% SECTION:  Beyond dipolar excitation                                                                            %%%
%%%%%%%%%%%%%%%%%%%%%%%%%%%%%%%%%%%%%%%%%%%%%

\subsection{Other excitations}
\vskip5mm

As an extension, we have studied the propagation of other collective excitations in the same system (for $d=10\mu_B$), in particular, the monopolar excitation (breathing mode) and quadrupolar ones ($2x^2-y^2-z^2$, $yz$, $xy$).

In order to get a good approximation of the frequency of a collective mode for a single dBEC, one can use the following equations obtained for contact interaction \cite{Dalfovo2000,Stringari1996}:
\begin{eqnarray} 
\omega^2 (m=0)&=&\omega_\perp^2 \left(2+\frac{3}{2}\beta^2\mp \frac{1}                   {2}\sqrt{9\beta^4-16\beta^2+16}\right)\label{mono}                                     \\
 \omega^2(m=\pm l)&=&l \, \omega_\perp^2 \label{quad1}                                        \\
\omega^2(m=\pm (l-1))&=&(l-1) \,\omega_\perp^2+\omega_x ^2 \,,\label{quad2}
\end{eqnarray}
where $\beta=\omega_x/\omega_\perp$ is the anisotropy of the BEC (in our case $\beta=30$), $l=1$ for center-of-mass modes, and $l=2$ for quadrupolar ones. The case $m=0$ refers the monopolar mode, which has to be dealed separatedly. Since we consider a pancake-shaped condensate with large $\beta$, we will see that not all these frequencies will be excited in our simulations neither even all these modes are totally decoupled. 

%For a dBEC the frequencies of the collective modes are slightly shifted with respect to the contact interacting ones (\ref{mono}-\ref{quad2})

%%%%%%%%%%%%%%%%%%%%%%%%%%%%%%%%%%%%%%%%%%%%%
%%% SUBSECTION: Monopolar excitations                                                         %%%
%%%%%%%%%%%%%%%%%%%%%%%%%%%%%%%%%%%%%%%%%%%%%

\subsubsection{Monopolar excitations.}
\vskip5mm

As regards the monopolar excitation, when $\beta\gg1$, from (\ref{mono}) there are two frequencies with $m=0$: $\omega_1=\sqrt{10/3}\,\omega_\perp\approx 1.83\,\omega_\perp$ and $\omega_2=\sqrt{3}\,\omega_x\,.$
%\begin{eqnarray}
%\omega_1&=&\sqrt{10/3}\,\omega_\perp\approx 1.83\,\omega_\perp \label{mono1} \\
%\omega_2&=&\sqrt{3}\,\omega_x\,.\label{mono2}
%\end{eqnarray}

Although these results have been obtained for contact interaction, in presence of dipolar interaction, the values are slightly 
shifted \cite{Lahaye2009,Bismut2010}. To quantify this shift, we have studied the monopolar response of a single dBEC with $\beta=30$, 
by solving the TDGPE and we have found  $\omega_1= 1.9 \,\omega_0$.

Now we study the propagation of the monopolar excitation between two non-overlapping dBECs. The perturbation corresponding to the monopolar excitation (breathing mode) is the following one (only applied on the left condensate):
\begin{equation}
\Psi(\vec r, 0) \to \Psi_{GS}(\vec r) \;e^{i\lambda (x^2+ y^2+ z ^2)}\,.
\label{monokick}
\end{equation}
Evolving this wavefunction in time by solving the TDGPE, we find similar results to the dipolar drag case, but with different frequencies (Fig. \ref{quadmodes}). In other words, only the low-lying energy frequency ($\omega_1$) and its corresponding shifted frequency are excited. In comparison with the dipolar drag effect, the magnitude of the shift in both cases is similar. Therefore, we see one frequency at $\omega=1.9\,\omega_\perp$ and its corresponding shifted frequency, showing that the coupled-pendulum model can be extended also for monopolar modes.

%%%%%%%%%%%%%%%%%%%%%%%%%%%%%%%%%%%%%%%%%%%%%
%%% SUBSECTION: Quadrupolar  excitations                                                     %%%
%%%%%%%%%%%%%%%%%%%%%%%%%%%%%%%%%%%%%%%%%%%%%

\subsubsection{Quadrupolar excitations.}
\vskip5mm

The frequencies corresponding to quadrupolar excitations in a pure contact interacting BEC are (\ref{quad1},\ref{quad2}): $\omega_1=\sqrt{2}\,\omega_\perp$ and $\omega_2=\,\omega_x\,$.
%\begin{eqnarray}
%\omega_1&=&\sqrt{2}\,\omega_\perp \label{quadr1} \\
%\omega_2&=&\,\omega_x\,.
%\end{eqnarray}
%
%{\color{red}
%TAULA AMB $\omega_{in}$ i $\omega_{out}$
%}
%
Three kind of quadrupolar excitations are implemented in the perturbation of the wavefunction, $2\,x^2-y^2-z^2$ (also known as quadrupolar), $xy$ and $yz$ respectively:
\begin{eqnarray}
\Psi(\vec r, 0) &\to& \Psi_{GS}(\vec r) \;e^{i\lambda (2\,x^2- y^2- z ^2)}\label{quadkick} \\
\Psi(\vec r, 0) &\to& \Psi_{GS}(\vec r) \;e^{i\lambda\, x\,y}\label{xykick} \\
\Psi(\vec r, 0) &\to& \Psi_{GS}(\vec r) \;e^{i\lambda \,y\,z}\,. \label{yzkick}
\end{eqnarray}

\begin{figure}[h!]
\begin{center}
\includegraphics[width=0.7\linewidth, clip=true]{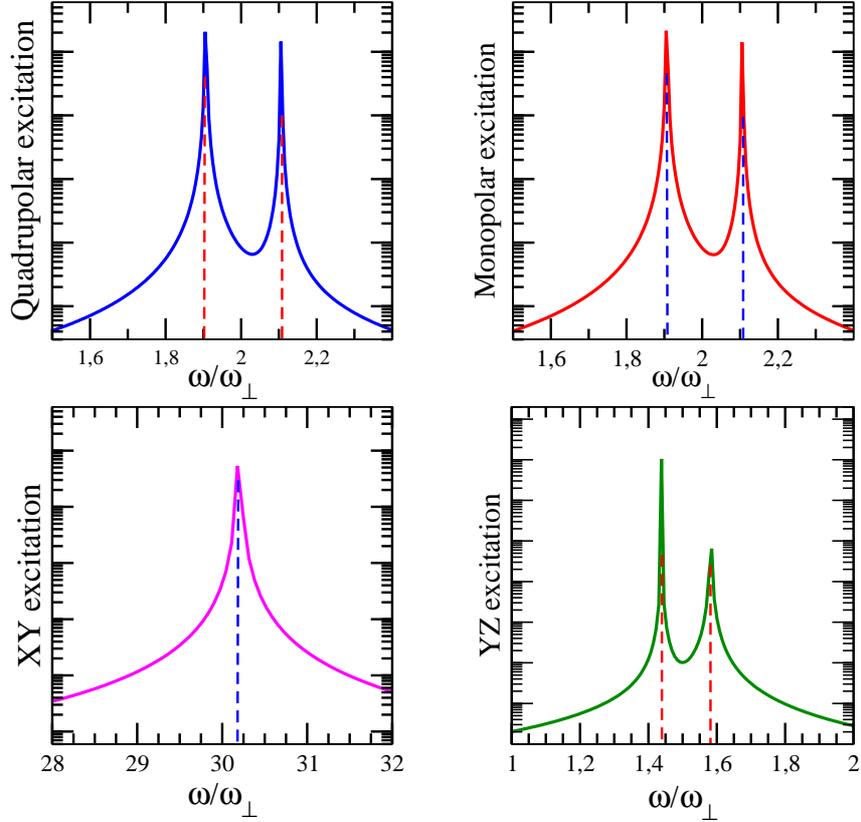}
\end{center}
\caption{\label{quadmodes} Fourier transform of the monopolar and quadrupolar 
responses ($2\,x^2- y^2-z^2$, $xy$ and $yz$). The strength is in logarithmic 
scale and arbitrary units.}
\end{figure}

Looking at the results of the time evolution (Fig. \ref{quadmodes}), those of the $2\,x^2-y^2-z^2$ excitation leads to a surprise, owing to the fact that it is shockingly similar to the results obtained for monopolar excitations. The reason for this similarity is that in a pancake geometry, with large $\beta$, the quadrupolar excitation degenerates to the monopolar one because the $x$-component of the perturbation gets almost inhibited due to the strong confinement. The frequencies obtained are the same for both cases. 

\begin{table}[h!]
\begin{center}
 \begin{tabular}{| c || c | c | c |}
\hline
Monopolar & $x^2+y^2+z^2$ & $\omega_{in}=1.89\,\omega_\perp$ & $\omega_{out}=2.11\,\omega_\perp$ \\ \hline
 & $-2x^2+y^2+z^2$ & $\omega_{in}=1.89\,\omega_\perp$ & $\omega_{out}=2.11\,\omega_\perp$ \\ \cline{2-4}
Quadrupolar & $xy$ & $\omega_{in}\approx\omega_x$ & $\omega_{out}=\omega_{in}$ \\ \cline{2-4}
 & $yz$ & $\omega_{in}=1.45\,\omega_\perp$ & $\omega_{out}=1.58\,\omega_\perp$ \\ 
 \hline
\end{tabular}
\caption{\label{table} Table that presents for the monopolar and each quadrupolar excitation the shape of the excitation and the value of the \textit{in-phase} and the \textit{out-of-phase} frequency found out in the Fourier transform.}
\end{center}
\end{table}

In contrast, it is not the case for the $xy$ and $yz$ excitation, as can be seen in Fig. \ref{quadmodes}. The former excites only the $\omega_2=\omega_x$ frequency, whereas the latter excites $\omega_1=\,1.45\,\omega_\perp$, and a second shifted frequency. The responsible of these differences is the kind of excitation. The $xy$ excitation only has one term, which depends on $x$. It means that the relevant frequency in the trap will be $\omega_x$. However, since the dBEC is tightly confined in the $x$-direction, the corresponding shift due to the dipolar coupling cannot be resolved.

As regards the $yz$ excitation, since the perturbation does not affect the $x$-direction, the frequency excited will be that of lower energy. In the numerical results we obtain one peak in the spectrum at $\omega_1=\,1.45\,\omega_\perp$, and, in addition, a second frequency due to the dipolar interaction will appear in this case because the frequency of the trap in the transversal direction is small. 

By symmetry and geometry, $xz$ excitation is analogous to the $xy$ excitation, and $y^2-z^2$ will be also analogous to the $yz$ excitation. In a pancake geometry, one can recover one or the other simply rotating the coordinate system.
%\vskip1cm

%%%%%%%%%%%%%%%%%%%%%%%%%%%%%%%%%%%%%%%%%%%%%
%%% SUBSECTION:   Dipolar drag effect in a triple-well configuration                %%%
%%%%%%%%%%%%%%%%%%%%%%%%%%%%%%%%%%%%%%%%%%%%%

\section{Dipolar drag effect in a triple-well configuration}
\vskip5mm
Let us extend the dipolar drag effect to the case, of a triple-well trap. Some studies have been carried out in
triple-well configurations related with dipolar effects~\cite{Peter2012}, phase diagrams~\cite{Lahaye2010} or 
symmetry breaking~\cite{Yan2012}. In the classical analogy of three coupled-pendulums (\ref{npend1}-\ref{npend3}), the eigenvalues of the problem are: $\omega_{in}=\omega_\perp$, $\omega_{out,1}=\omega_\perp\sqrt{1+\alpha/\omega_\perp^2}$ and $\omega_{out,2}=\omega_\perp\sqrt{1+3\alpha/\omega_\perp^2}$.
%\begin{eqnarray}
%\omega_{in}&=&\omega_\perp\label{omegain3}                                                          \\
%\omega_{out,1}&=&\omega_\perp\sqrt{1+\alpha/\omega_\perp^2}\label{omegaout13}     \\
%\omega_{out,2}&=&\omega_\perp\sqrt{1+3\alpha/\omega_\perp^2}\,.\label{omegaout23}
%\end{eqnarray}

\begin{figure}[h!]
\begin{center}
\includegraphics[width=0.55\linewidth, clip=true]{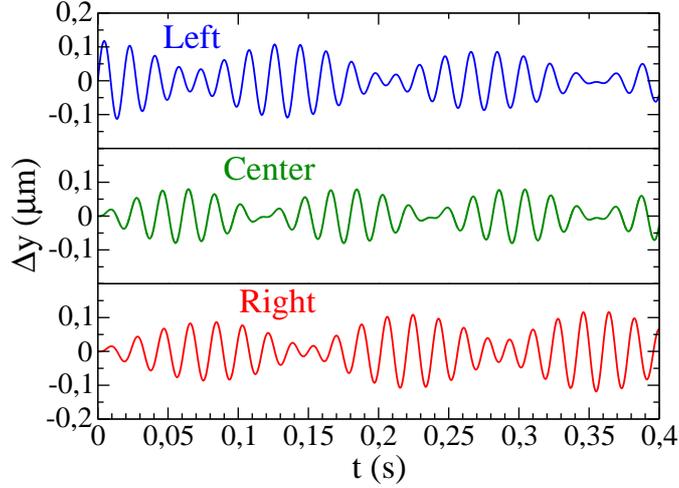}
\end{center}
\caption{\label{triplewellnosym}
Displacements of the centers of mass of the three condensates in a triple-well configuration when we excite only the 
left condensate. From top to bottom, condensate of the left, center and right.}
\end{figure}

In Fig. \ref{triplewellnosym}, we show the displacements of the centers of mass when only the left condensate is perturbed. All three frequencies appear in the movement, and, as a consequence, the displacements of the centers of mass have a more complex behaviour.

Although there are three eigenmodes, it is possible to excite only two frequencies by choosing the appropiate initial conditions. If we perturb only the condensate of the center, looking to the eigenvectors of the problem, only the two frequencies that preserve the symmetry of the system, $\omega_{in}$ and $\omega_{out,2}$, will appear in the movement of the center of mass (the other frequency corresponds to a non-symmetric mode).

\begin{figure}[h!]
\begin{center}
\includegraphics[width=0.55\linewidth, clip=true]{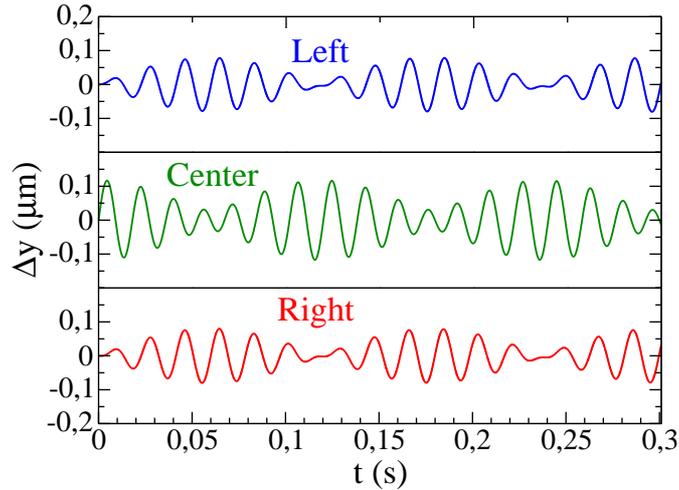}
\end{center}
\caption{\label{triplewell}
Displacements of the centers of mass of the three different condensates in the triple-well configuration, when we perturb only the center condensate. From top to bottom, condensate of the left, center and right.}
\end{figure}

In Fig. \ref{triplewell}, we present the numerical results obtained by solving the TDGPE exciting only the center of mass mode of the centered dBEC. As expected, only two frequencies are involved in the movement of the three non-overlapping condensates. We have obtained the corresponding frequencies by performing a Fourier analysis and we have compared them with the case of only two dBECs:
\begin{equation}
\frac{(\omega_{out,2} ^2-\omega_\perp ^2)_{3well}}{(\omega_{out} ^2-\omega_\perp ^2)_{2well}}\approx1.6\,.
\end{equation}
 It is very close to the value of the ratio $1.5$ obtained in the classical analogy. This means that the classical coupled-pendulum model yields a qualitative description of the dipolar coupling between few non-overlapping dBECs. Another remarkable fact is that, indeed, the magnitude of the shift is increased for three condensates compared with the case of two.

%%%%%%%%%%%%%%%%%%%%%%%%%%%%%%%%%%%%%%%%%%%%%
%%% SECTION:  Summary and conclusions                                                                          %%%
%%%%%%%%%%%%%%%%%%%%%%%%%%%%%%%%%%%%%%%%%%%%%

\section{Summary and conclusions}
\vskip5mm

In the present work, the full TDGPE has been numerically solved to study the long-range character of the dipolar interaction. 
We have investigated the propagation of collective modes, such as dipolar and quadrupolar excitations, between non-overlapping 
pancake-shaped dBECs.
A classical coupled-pendulum model has been proposed.

We have analyzed the effect of different parameters in the propagation of the excitations. The dipole moment enhances the dipolar 
effects, 
whereas the enlargement of the scattering length and the distance between condensates leads the system to be dominated by the contact 
interaction which hides the desired long-range effects.

%%%%%%%%%%%%%%%%%%%%%%%%%%%%%%%%%%%%%%%%%%%%%
%%% REFERENCES                                                                                             %%%
%%%%%%%%%%%%%%%%%%%%%%%%%%%%%%%%%%%%%%%%%%%%%

%
%\bibitem{Peter2012}
%D. Peter, \textit{et al}, J. Phys. B: At. Mol. Opt. Phys. {\bf 45}, 225302 (2012).


\section*{References}
\vskip5mm
\begin{thebibliography}{99}
\bibitem{Griesmaier2005} 
Griesmaier A, Werner J, Hensler S, Stuhler J and Pfau T 2005 
{\it Phys. Rev. Lett}. {\bf 94} 160401 
%
\bibitem{Lu2011} 
Lu M, Burdick  N  Q , Youn S H and Lev B L  2011 
{\it Phys. Rev. Lett}. {\bf107} 190401 
%
\bibitem{Aikawa2012}
Aikawa K, Frisch A, Mark M, Baier S, Rietzler A, Grimm R and Ferlaino F 2012
{\it Phys. Rev. Lett}. {\bf108} 210401 
\bibitem{Lahaye2009} 
Lahaye T, Menotti C, Santos L, Lewenstein M and Pfau T 2009 
{\it Rep. Prog. Phys}. {\bf72} 126401
%
\bibitem{Abad2010} 
Abad M, Guilleumas M, Mayol R, Pi M and Jezek D M 2010 
{\it Phys. Rev. A}. {\bf 81} 043619
%
\bibitem{Lahaye2010} 
Lahaye T, Pfau T and Santos L 2010
Phys. Rev. Lett.  {\bf 104} 170404
%
\bibitem{Yan2012} 
Yan P, Wang Y, Ji S and Liu X 2012
{\it Phys. Lett. A}. {\bf 376} 3141
%
\bibitem{Matveeva2011}
Matveeva N, Recati A and Stringari S 2011 
{\it Eur. Phys. J. D}.  {\bf 65} 219 
%
\bibitem{Huang2010} 
Huang C and Wu W 2010 
{\it Phys. Rev. A}. {\bf 82} 053612
%
\bibitem{Bismut2010} 
Bismut G, Pasquiou B, Mar\'echal E, Pedri P, Vernac L, Gorceix O and Laburthe-Tolra B 2010 {\it Phys. Rev. Lett}. {\bf 105} 040404
%
\bibitem{Hu1996} 
Hu G Y and O'Connell R F 1996 
{\it J. Phys. A: Math}. {\bf 29} 1511
%
\bibitem{Jauho1993} 
Jauho A  P and Smith H 1993
Phys. Rev. B. {\bf 47} 4420
%
\bibitem{Rojo1999} 
Rojo A G 1999 
{\it J. Phys.: Condens. Matter}. {\bf 11} R31
%
\bibitem{Abad2011} 
Abad M, Guilleumas M, Mayol R, Pi M and Jezek D M 2011 {\it Phys. Rev. A}. {\bf 84} 035601
%
\bibitem{Calvayrac1997} 
Calvayrac F, Reinhard P G, Suraud E 1997 
{\it Ann. Phys}. {\bf 255} 125
%
\bibitem{Dalfovo2000}
Dalfovo F, Giorgini S, Pitaevskii L and Stringari S 1999 
{\it Rev. Mod. Phys}. {\bf 71} 463
%
\bibitem{Stringari1996}
Stringari S 1996
{\it Phys. Rev. Lett}. {\bf 77} 2360

%
\bibitem{Peter2012}
Peter D, Pawlowski K, Pfau T and  Rzazewski K 2012
{\it  J. Phys. B: At. Mol. Opt. Phys}. {\bf 45} 225302 
\end{thebibliography}
\end{document}